\documentclass[a4paper]{jpconf}
\usepackage{amssymb,amsmath,graphicx}
\begin{document}
\newcommand{\be}{\begin{equation}}
\newcommand{\ee}{\end{equation}}
\newcommand{\bea}{\begin{eqnarray}}
\newcommand{\eea}{\end{eqnarray}}
\newcommand{\beas}{\begin{eqnarray*}}
\newcommand{\eeas}{\end{eqnarray*}}

\def \CMP {{\it Commun. Math. Phys.}}
\def \PRL {{\it Phys. Rev. Lett.}}
\def \PL {{\it Phys. Lett.}}
\def \NPBProc {{Nucl. Phys. B (Proc. Suppl.)}}
\def \NP {{\it Nucl. Phys.}}
\def \RMP {{Rev. Mod. Phys.}}
\def \JGP {{J. Geom. Phys.}}
\def \CQG {{Class. Quant. Grav.}}
\def \MPL {{Mod. Phys. Lett.}}
\def \IJMP {{ Int. J. Mod. Phys.}}
\def \JHEP {{\it JHEP}}
\def \PR {{\it Phys. Rev.}}
\def\del  {{\partial}}

\title{Casimir effect: Edges and diffraction}

\author{Dimitra Karabali}

\address{Department of Physics and Astronomy, Lehman College of the CUNY, Bronx, NY 10468, USA}

\ead{dimitra.karabali@lehman.cuny.edu}

\begin{abstract}
The Casimir effect refers to the existence of a macroscopic force between conducting plates in vacuum due to quantum fluctuations of fields. These forces play an important role, among other things, in the design of nano-scale mechanical devices. Accurate experimental observations of this phenomenon have motivated the development of new theoretical approaches in dealing with the effects of different geometries, temperature etc. In this talk, I will focus on a new method we have developed in calculating the contribution to the Casimir effect due to diffraction from edges and holes in different geometries, at zero and at finite temperature. 
\end{abstract}

\section{Introduction}
Casimir effect displays in a beautiful way the emergence of a macroscopic force due to quantum fluctuations in vacuum. The original Casimir effect \cite{Casimir} described the interaction between two parallel, infinitely long conducting plates. It was shown that due to electromagnetic fluctuations in vacuum, there is an attractive force between the plates given by $f= - \partial E / \partial a$, where
\be
E = - {\pi^2 A \over {720 a^3}}
\label{energyinf}
\ee
$A$ is the area of the plates and $a$ is the distance between them. One way to think about this is that the presence of plates imposes boundary conditions which modify the field modes. As a result, the zero-point energy contribution of the fields gets shifted. This is of course infinite, but the difference in zero-point energy with and without the plates is finite and produces (\ref{energyinf}).

There has been a renewed interest in the Casimir effect over the last few years \cite{reviews}, driven by the fact that: a) advances in instrumentation have allowed precise measurement of the effect \cite{Lamoreaux} and b) the Casimir force becomes relevant in the design of nano-scale mechanical devices where it can cause the tiny elements in such devices to stick together. This has spurred a lot of activity on the theory side in terms of developing approaches to deal with the effects of different geometries, orientation, surface roughness, thermal fluctuations etc., issues relevant in realistic experimental setups. 

An important and rather difficult question to address is how diffractive effects correct the Casimir energy. This is relevant whenever the plates have boundaries, either apertures or edges. Edge contributions have been studied numerically in a few cases using worldline techniques and Monte Carlo simulations \cite{Gies}-\cite{GiesWeber}. A systematic treatment of diffraction is lacking in previous analytical methods of the Casimir effect. In this talk I will describe our recent work towards understanding such diffractive effects at zero and finite temperature. 

This talk is based on work done in collaboration with D. Kabat and V.P. Nair \cite{Kabat1}-\cite{Kabat3}. I will first describe the general analytical approach we have developed to study diffractive effects in the context of Casimir effect and then apply this to different geometries such as a single plate with a slit in it, two perpendicular plates separated by a gap, and two parallel plates, one of which is semi-infinite, at zero and finite temperature. Other approaches to analyzing the Casimir energy in such geometries are given in \cite{MIT}. An advantage of our formalism is that it allows for a clean separation between direct or geometrical effects associated with the plates,
and diffractive effects associated with the plate boundaries. Finally, I will describe a novel application of our formalism in studying the Casimir interaction between far away holes or slits on a flat Dirichlet plate.

\section{An effective boundary action for edge effects\label{sec:general}}

For simplicity we consider a free massless scalar field in four Euclidean dimensions, with Dirichlet boundary
conditions imposed on an arrangement of plates. The basic plate geometry we will consider is shown in Fig.1. The field propagates in two regions separated by a plate with a gap, indicated by a dashed line. The field vanishes everywhere on the boundary (solid lines), while $\phi = \phi _0$ on the gap. 
\begin{figure}
\begin{center}
\scalebox{.8}{\includegraphics{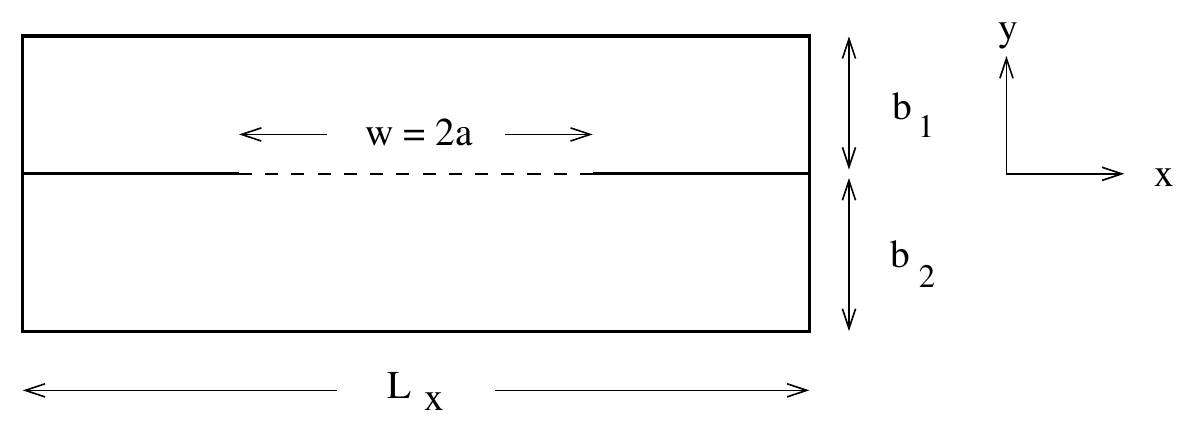}}
\end{center}
\caption{A 2d slice of the full geometry.  
The 4d geometry also has a periodic spatial dimension of
size $L$ out of the page and a periodic Euclidean time dimension of size $\beta$. We eventually take the limit $L_x,~L \rightarrow \infty$.
\label{fig:FullGeometry}}
\end{figure}
We will follow a path integral approach and calculate the free energy $F$ in terms of the partition function 
\be
F = -{1\over \beta} \, \log Z 
\ee 
where
\be
Z = \int [d\phi] ~e^{- S (\phi )}, \hskip .2in
S (\phi ) = {1\over 2}\int_0^\beta d\tau d^3x \, \, (\del \phi \, \,\del \phi )
\label{Z}
\ee
The basic strategy is to calculate the partition function in stages. We first fix the value of the field in the gap, $\phi |_{\rm gap} = \phi _0$, integrate out the scalar field in the bulk top and bottom regions and eventually integrate over $\phi_0$. By integrating out the scalar field in the bulk regions we obtain a lower dimensional, non-local effective action in the gap in terms of $\phi_0$ as follows. To perform the bulk path integral we set
$\phi = \phi_{\rm cl} + \delta \phi$
where $\delta \phi$ vanishes on all boundaries (including the gap), and $\Box \phi_{\rm cl} = 0$, subject to the boundary conditions
\be
\phi_{\rm cl} \rightarrow \left\lbrace \begin{array}{ll}
\phi_0 & \hbox{\rm in gap} \\
0 & \hbox{\rm elsewhere on boundary}
\end{array} \right.
\label{pathint4}
\ee
$\phi_{\rm cl}$ can be written in terms of $\phi_0$ and the Green's functions $G_{\rm top}$ and
$G_{\rm bottom}$.  These obey Dirichlet
boundary conditions and satisfy
$\Box G(x \vert x') = \delta^4(x-x')$ in the bulk regions. 
\begin{equation}
\label{Greens}
\phi_{\rm cl}(x) = \left\lbrace \begin{array}{ll}
\int d^3x' \, \phi_0(x') \, n \cdot \partial' G_{\rm top}(x \vert x') & \hbox{\rm on top} \\[8pt]
\int d^3x' \, \phi_0(x') \, n \cdot \partial' G_{\rm bottom}(x \vert x') & \hbox{\rm on bottom}
\end{array}\right.
\end{equation}
where $n$ is an outward-pointing unit normal vector.  Integrating by
parts, the original action in (\ref{Z}) becomes,
\bea
S (\phi )& = & {1\over 2}\int (\del \eta \, \del \eta )_L + 
{1\over 2}\int (\del \eta \, \del \eta )_R +  S_0 \nonumber \\
S_0 &=&  \int_{gap} ~ {1 \over 2} \phi_0(x) \, (M_{\rm top} (x\vert x') + M_{\rm bottom}(x\vert x'))~ \phi_0 (x')  
\label{S0}
\eea
where 
\be
M (x \vert x') = n\cdot \del \,~ n \cdot \del' \, G(x\vert x')_{L,R}
\ee
We can write a mode expansion for the fields
$\phi_0$ as
$\phi_0 ({x})
= \sum_\alpha c_\alpha u_\alpha ({ x})$, where $\left\{ u_\alpha ({x})
\right\}$ constitute a complete set of modes for functions which are nonzero
in the gap with the boundary condition that $u_\alpha ({x})\rightarrow 0$
as one approaches the edges. Integrating over $c_\alpha$ we get
\be
\label{Z4d}
Z_{\rm 4d} = \det{}^{-1/2}\big(-\Box_{\rm top}\big) \,
            \det{}^{-1/2}\big(-\Box_{\rm bottom}\big) \,
            \det{}^{-1/2}\big({\cal O}_{\rm top} + {\cal O}_{\rm bottom}\big)
\ee
where
\be
\label{matrix elements}
{\cal O}_{\alpha\beta} = \int_{\rm gap} d^{3}{x}~d^{3}{x'}~ u_\alpha (x) M (x\vert x') ~ u_\beta (x')
\ee
The bulk determinants in (\ref{Z4d}) capture the Casimir energy that would be present if there was no gap in the middle plate. Corrections to this are given by the non-local field theory $S_0$ in (\ref{S0}) that lives on the gap separating the two regions. 

The explicit form of the operator ${\cal O}$ depends, in general,
on the arrangement of plates and gaps. For the geometry shown in Fig.1 we find
\bea
M_R (x\vert x' )
&=& <x| {\sqrt{-\nabla^2}  \coth \big(b_1 \sqrt{-\nabla^2}\big)} |x'> \nonumber\\
M_L (x\vert x') &=& <x| {\sqrt{-\nabla^2}  \coth \big(b_2 \sqrt{-\nabla^2}\big)} |x'>  \nonumber
\eea
where $\nabla^2$ is the Laplacian on the middle plate. Further it is convenient to make a Kaluza-Klein decomposition along the two extra periodic directions.  This leads to a
representation of the four dimensional partition function in terms of a momentum integral and a sum over Matsubara frequencies.
\be
\label{Z4dKK}
\log Z_{\rm 4d} = L \int {dk \over 2\pi} \sum_{l = -\infty}^\infty
\log Z_{\rm 2d}\Big(\mu = \sqrt{k^2 + (2 \pi l / \beta)^2}\Big)
\ee
Here $Z_{\rm 2d}(\mu)$ is the two-dimensional partition function for a
scalar field of mass $\mu$ in the geometry shown in
Fig.1.

For the geometry of Fig.~\ref{fig:FullGeometry}, a complete set of  odd- and even-parity functions which vanish for $\vert x \vert \geqslant  a$ are
\bea
\label{OddModeFtns}
&& u^{\rm odd}_m = \left\lbrace
\begin{array}{cl}
(-1)^m {1\over \sqrt{a}} \sin \left(m \pi x / a\right) & \hbox{\rm for $-a \leqslant  x \leqslant a$}  \qquad m = 1,2,3,\ldots\\
0 & \hbox{\rm otherwise}
\end{array}\right.  \\
\label{EvenModeFtns}
&& u^{\rm even}_p = \left\lbrace
\begin{array}{cl}
(-1)^{p + {1\over 2}} {1\over \sqrt{a}} \cos \left(p \pi x / a\right) & \hbox{\rm for $-a \leqslant  x \leqslant a$}  \qquad p = {1 \over 2},{3 \over 2},{5 \over 2},\ldots\\
0 & \hbox{\rm otherwise}
\end{array}\right. 
\eea
The matrix elements of the operator ${\cal O}$ in this basis are
\bea
\label{OddMatrix}
{\cal O}^{\rm odd}_{mn} & = & {2 a \over \pi} \int_{-\infty}^\infty dk \, \sin^2(ka) \, M(k) \, {m\pi \over k^2 a^2 - m^2 \pi^2} \,
{n\pi \over k^2 a^2 - n^2 \pi^2} \\
{\cal O}^{\rm even}_{pq} & = & {2 a \over \pi} \int_{-\infty}^\infty dk \, \cos^2(ka) \, M(k) \, {p\pi \over k^2 a^2 - p^2 \pi^2} \,
{q\pi \over k^2 a^2 - q^2 \pi^2} 
\label{integrals}
\eea
where $m,n=1,2,\cdots$,~$p,q=1/2,3/2,\cdots$, and  $M(k) = {\sqrt{k^2 + \mu^2} \over \tanh \big(b \sqrt{k^2 + \mu^2}\big)}$ has the useful representation
\[
M(k) = {1 \over b} + {2 \over b} \sum_{j = 1}^\infty {k^2 + \mu^2 \over k^2 + \mu^2 + {j^2 \pi^2 \over b^2}}\,.
\]
The integrals (\ref{OddMatrix}), (\ref{integrals}) are evaluated using a contour deformation and they separate naturally into  a pole contribution, ${\cal O}^{\rm direct}$, and a cut contribution, ${\cal O}^{\rm diffractive}$. (In \cite{Kabat1} these were referred to
as ``pole'' and ``cut'' contributions, respectively.)  For the odd matrix elements
\bea
\label{Odd}
&& {\cal O}^{\rm odd}_{mn} = {\cal O}_{mn}^{\rm direct} + {\cal O}_{mn}^{\rm diffractive} \nonumber  \\
\label{OddDirect}
&& {\cal O}_{mn}^{\rm direct} = {\sqrt{(m \pi / {a})^2 + \mu^2} \over \tanh \big(b \sqrt{(m \pi / {a})^2 + \mu^2}\big)} \, \delta_{mn} \\
\nonumber
&& {\cal O}_{mn}^{\rm diffractive} = - 2 a b^2 \sum_{j = 1}^\infty
\left(1 - \exp\Big(-{2 a \over b}\sqrt{j^2 \pi^2 + \mu^2 b^2}\Big)\right)
{j^2 \pi^2 \over \sqrt{j^2 \pi^2 + \mu^2 b^2}} \\
\label{OddDiffractive}
&& \qquad \qquad \qquad {m \pi \over (m \pi b)^2 + (j \pi a)^2 + (\mu a b)^2} \,\,
                       {n \pi \over (n \pi b)^2 + (j \pi a)^2 + (\mu a b)^2}
\eea
Likewise for the even matrix elements
\bea
\label{Even}
&& {\cal O}^{\rm even}_{pq} = {\cal O}_{pq}^{\rm direct} + {\cal O}_{pq}^{\rm diffractive} \nonumber  \\
\label{EvenDirect}
&& {\cal O}_{pq}^{\rm direct} = {\sqrt{(p \pi / {a})^2 + \mu^2} \over \tanh \big(b \sqrt{(p \pi / {a})^2 + \mu^2}\big)} \, \delta_{pq} \\
\nonumber
&&{\cal O}_{pq}^{\rm diffractive} = - 2 a b^2 \sum_{j = 1}^\infty
\left(1 + \exp\Big(-{2 a \over b}\sqrt{j^2 \pi^2 + \mu^2 b^2}\Big)\right)
{j^2 \pi^2 \over \sqrt{j^2 \pi^2 + \mu^2 b^2}} \\
\label{EvenDiffractive}
&& \qquad \qquad \qquad {p \pi \over (p \pi b)^2 + (j \pi a)^2 + (\mu a b)^2} \,\,
                       {q \pi \over (q \pi b)^2 + (j \pi a)^2 + (\mu a b)^2}
\eea
(Aside from the allowed values of the indices, the only difference between odd and even parity is the sign
in front of the exponential in the diffractive term.)  

The direct contribution takes into account the geometric optics effect of wave propagation directly across the gap.  ${\cal O}^{\rm direct}$ is essentially the operator $M$ defined on the gap with Dirichlet boundary conditions at $x= \pm a$.  Corrections to this, which incorporate diffraction of
waves through the gap, are encoded in ${\cal O}^{\rm diffractive}$.

The strategy now is to treat diffraction as a perturbation and expand in powers of diffractive contributions. 
Taking the log of (\ref{Z4d}) and expanding
in powers of ${\cal O}^{\rm diffractive}$, the free energy naturally
decomposes into bulk, direct and diffractive contributions.
\bea
\label{bulk}
-\log Z_{\rm bulk} &=& {1 \over 2} {\rm Tr} \log \big(-\Box_{\rm top}\big)
                              + {1 \over 2} {\rm Tr} \log \big(-\Box_{\rm bottom}\big) \\
\label{direct}
-\log Z_{\rm direct} &=& {1 \over 2} {\rm Tr} \log \big({\cal O}^{\rm direct}\big) = {1 \over 2} {\rm Tr} \log \big({\cal O}_{\rm top}^{\rm direct}
                                                               + {\cal O}_{\rm bottom}^{\rm direct}\big) \\
\label{diffractive}
-\log Z_{\rm diffractive} &= &{1 \over 2} {\rm Tr} \left[ 
\left({\cal O}^{\rm direct}\right)^{-1} {\cal O}^{\rm diffractive} \right] \nonumber \\
\nonumber
&-&{1 \over 4} {\rm Tr }\left[ 
\left({\cal O}^{\rm direct}\right)^{-1}
{\cal O}^{\rm diffractive}
\left({\cal O}^{\rm direct}\right)^{-1}
{\cal O}^{\rm diffractive} \right] + \cdots\\[3pt]
\label{expansion}
&=& -\log Z^{(1)}_{\rm diffractive} -\log Z^{(2)}_{\rm diffractive} + \cdots
\eea

We have applied this approach to derive the Casimir energy in the case of three special geometries: a single plate with a slit, perpendicular plates and parallel plates, where one is infinite and the other semi-infinite or with a large gap. This is done by considering appropriate limits of the geometric parameters $b_1, b_2$ and $a$ in Fig. 1. Further by taking different limits of $\beta$ we can calculate the Casimir energy at various temperature regimes. In the next two sections I will give a brief summary of our results in these cases.

\section{Zero temperature Casimir energy\label{sec:zeroT}}

In the zero-temperature case, $\beta \rightarrow \infty$, the sum over Matsubara modes in (\ref{Z4dKK}) becomes an integral and the free energy is
\be
F = -{1 \over \beta} \log Z{}_{4d} = L~ \int_0^\infty {\mu d\mu \over 2 \pi} \, \left(- \log Z{}_{2d}\right)\,.
\label{slit-diff15}
\ee
In evaluating the free energy, we encounter ultraviolet divergencies. In general, all such terms are proportional to geometrical volumes, areas, perimeters, etc and once they are appropriately renormalized, they are not part of the Casimir energy. 

\subsection{Plate with a slit}

We first consider a plate with a single slit of width $w=2a$ and length $L$. This corresponds to $b_1, b_2 \rightarrow \infty$. There are direct and diffractive contributions to the Casimir energy: 
\bea
F^{\rm direct}_{{\rm slit}} &=& -{{\zeta(3) L} \over {32 \pi w^2}} =  - 11.96 \times 10^{-3} ~ {L\over w^2} \nonumber\\
F^{(1)\rm diffractive}_{{\rm slit}} &=&  8.60  \times 10^{-3}~ {L\over w^2}, \hskip .35in
F^{(2)\rm diffractive}_{{\rm slit}} =  0.56  \times 10^{-3} ~{L\over w^2} \nonumber\\
F^{(3)\rm diffractive}_{{\rm slit}} &=&  0.08  \times 10^{-3} ~{L\over w^2}, \hskip .35in
F^{(4)\rm diffractive}_{{\rm slit}} = 0.02  \times 10^{-3}~{L\over w^2} \label{slit-diff18}
\eea
The total value for the Casimir energy up to this order is $F_{{\rm slit}}= -2.70\times 10^{-3} (L/w^2)$. 

\subsection{Perpendicular plates}

For perpendicular plates distance $a$ away (and  $b_1,b_2 \rightarrow \infty$), we only need to keep the odd-parity modes (\ref{OddModeFtns}) and matrix elements (\ref{OddMatrix}) in calculating the direct and diffractive contributions to the Casimir energy. We find
\bea
F^{\rm direct}_{\perp} &=&  -{{\zeta(3) L} \over {32 \pi a^2}} =  - 11.96 \times 10^{-3}~{L\over a^2} \nonumber\\
F^{(1)\rm diffractive}_{\perp}  &=&  5.01  \times 10^{-3}~{L\over a^2}, \hskip .35in
F^{(2)\rm diffractive}_{\perp}  = 0.66  \times 10^{-3}~{L\over a^2}\nonumber\\
F^{(3)\rm diffractive}_{\perp}  &=& 0.16  \times 10^{-3}~{L\over a^2}, \hskip .35in
F^{(4)\rm diffractive}_{\perp}  = 0.05  \times 10^{-3}~{L\over a^2}\label{slit-diff18}\\
F^{(5)\rm diffractive}_{\perp}  &=& 0.01  \times 10^{-3}~{L\over a^2}\nonumber
\eea
The total value for the Casimir energy up to this order is $F_{\perp}=-6.07\times 10^{-3} (L/a^2)$. This is in very good agreement with the worldline results of Gies and Klingm\"uller \cite{Gies} and analytical results derived by the multiple scattering method \cite{MIT}. 

\subsection{Infinite, semi-infinite parallel plates}

In the case of two long parallel plates, distance $b$ away, where one has a large slit ($a, b_1 \rightarrow \infty, b_2 \rightarrow b$) there is a finite bulk contribution from the bottom bulk region in Fig.1 in addition to the direct and diffractive contributions. 


\be
\label{bulk energy}
F^{\rm bottom~bulk}_{||} = -{\pi^2 \over {1440 b^3}}L_x L + \cdots
\ee
The $\cdots$ indicate possible edge terms at infinity (associated with the walls of the box shown in Fig.1). These terms depend on the boundary conditions at infinity and do not affect the slit contribution. 

The direct contribution to the energy as $a \rightarrow \infty$ is
\be
\label{direct energy}
F^{\rm direct}_{||} = {\pi^2 \over {1440 b^3}}2 a L - {\zeta(3) \over {32 \pi b^2}} L
\ee
The total bulk contribution is proportional to the area $A$ facing the two plates 
\be
\label{bulkanddirect}
F^{\rm bulk}_{||} =-{{\pi^2 L(L_1-2a)} \over {1440 b^3}} =-{{\pi^2A} \over {1440 b^3}} 
\ee
The diffractive contribution up to the 5th order are,
\bea
F^{(1)\rm diffractive}_{||} &=&  5.54 \times 10^{-3}~{L \over b^2}, \hskip .2in
F^{(2)\rm diffractive}_{||}  =  0.80\times 10^{-3}~{L \over b^2}\nonumber\\
F^{(3)\rm diffractive}_{||}  &=&  0.19 \times 10^{-3}~{L \over b^2}, \hskip .27in
F^{(4)\rm diffractive}_{||}  =  0.05  \times 10^{-3}~{L \over b^2}\nonumber\\
F^{(5)\rm diffractive}_{||}  &=&  0.01 \times 10^{-3}~{L \over b^2}\label{||plate17}
\eea
We find from (\ref{direct energy}) an (\ref{||plate17}) that the total edge contribution to the Casimir energy, up to this order  is 
\be
F^{\rm edge}_{||} = -{{\zeta(3) L} \over {32 \pi b^2}} + F^{\rm diffractive}_{||} \sim -5.37 \times 10^{-3} {L \over b^2}
\label{edge}
\ee
In our case there are two edges associated with the large slit. In comparing our results to the case of two parallel plates, one of which is semi-infinite, where there is only one slit edge, we have to divide  (\ref{edge}) by a factor of two. The result is in very good agreement with the worldline results of Gies and Klingm\"uller \cite{Gies} and those obtained using the multiple scattering method \cite{MIT}. 

Before I continue to discuss thermal effects, I would like to highlight two features of the diffractive terms in all the geometries discussed earlier. First, the series expansion in terms of diffractive contributions seems to converge nicely, justifying the perturbative expansion idea. Second, all diffractive energy terms correspond to an effective repulsive force between the plates, opposite to the attractive effect arising from the direct contribution. In all cases we considered though, the diffractive effects are subdominant compared to the direct ones, leading to an overall attractive edge Casimir force. It is also interesting to note that the diffractive contribution in the case of the slit is much bigger than the one for the perpendicular plates, as expected in physical grounds. These features remain the same in the case of finite temperature as well. 

\section{Thermal effects: high and low-temperature limits\label{sec:temp}}
\subsection{High temperature limit}

An important aspect of the Casimir effect is its dependence on temperature. At high temperature the free energy has a universal dependence on $T$ independent of the geometry of the plates. This can be seen from (\ref{Z4dKK}), where, in the limit  $\beta \rightarrow 0$ only the $l=0$ mode contributes and the problem reduces to a partition function in three dimensions which is independent of $T$. Below I will briefly present our results for the Casimir energy at high temperature for the three geometries considered earlier.
\vskip .15in
\noindent
{\it Plate with a slit}

The direct contribution to the thermal Casimir energy for a slit of width $w$, as $wT \rightarrow \infty$, is
\be
F^{\rm direct}_{\rm slit,T} = -{{\zeta(2)LT} \over {8 \pi w}} 
\label{slithighT}
\ee
The first four diffractive contributions are:
\bea
F^{(1)\rm diffractive}_{\rm slit,T} &=&  0.03901~{LT \over w}, \hskip .2in
F^{(2)\rm diffractive}_{\rm slit,T}  =  0.00431~{LT \over w}\nonumber\\
F^{(3)\rm diffractive}_{\rm slit,T}  &=&  0.00092~{LT \over w}, \hskip .27in
F^{(4)\rm diffractive}_{\rm slit,T}  =  0.00027~{LT \over w}\nonumber
\eea
The $w$-dependent part of the thermal Casimir energy up to this order is
\be
F_{\rm slit, T} = -0.02094~ {LT \over w}
\ee
\vskip .15in
\noindent
{\it Perpendicular plates}

The direct contribution to the thermal Casimir energy for two perpendicular plates distance $a$ apart is the same as in (\ref{slithighT}), where $w$ is replaced by $a$. The first four diffractive contributions are:
\bea
F^{(1)\rm diffractive}_{\perp,T} &=&  0.02161~{LT \over a}, \hskip .2in
F^{(2)\rm diffractive}_{\perp,T}  =  0.00320~{LT \over a}\nonumber\\
F^{(3)\rm diffractive}_{\perp,T}  &=&  0.00082~{LT \over a}, \hskip .27in
F^{(4)\rm diffractive}_{\perp,T}  =  0.00025~{LT \over a}\nonumber
\eea
The $a$-dependent part of the total thermal Casimir energy up to this order is
\be
F_{\perp, T} = -0.03957 {LT \over w}
\ee
This result agrees very well with the worldline results in \cite{GiesWeber}. 
\vskip .15in
\noindent
{\it Infinite, semi-infinite parallel plates}

The renormalized bulk and direct contributions to the thermal Casimir energy is
\bea
F^{\rm bulk}_{||,T} &=& - {{A \zeta(3)} \over {16 \pi b^2}}T + \cdots \nonumber \\
F^{\rm direct}_{||,T} &=& -{{\zeta(2) LT} \over {16 \pi b}} 
\eea
where $A$ is the area of the semi-infinite plate and $\cdots$ indicate edge terms associated with the boundaries at infinity. The first four diffractive contributions are:
\bea
F^{(1)\rm diffractive}_{||,T} &=&  0.01351~{LT \over b}, \hskip .2in
F^{(2)\rm diffractive}_{||,T}  =  0.00225~{LT \over b}\nonumber\\
F^{(3)\rm diffractive}_{||,T}  &=&  0.00056~{LT \over b}, \hskip .27in
F^{(4)\rm diffractive}_{||,T}  =  0.00015~{LT \over b}\nonumber
\eea
The total edge contribution to the thermal Casimir energy up to this order is
\be
F^{\rm edge}_{||,T} = -0.016126~{LT \over b}
\ee
which is again in excellent agreement with the results in \cite{GiesWeber}.

\subsection{Low temperature limit}

The behavior at low temperature is more subtle. In particular, at low temperature, thermal effects are dominated by long-range fluctuations which are suppressed in closed geometries but not so in open geometries. One then expects a non-trivial correlation between geometry and temperature. This was studied numerically using the worldline formalism in a number of geometries \cite{KlingGies}, \cite{GiesWeber}. Our formalism on the other hand provides an analytic way to understand the temperature-geometry interplay and in particular the role diffraction plays \cite{Kabat3}.

In the special geometries we studied earlier the free energy has three contributions: the bulk, the direct and the diffractive contributions. The bulk contribution from the regions above and below the middle plate is that of a 4d ideal Bose gas, while the direct contribution is that of a 3d ideal Bose gas. There are explicit analytic expressions for these, and the high and low temperature limits are easy to derive (see for example Appendix in \cite{Kabat3}). The diffractive contribution however, is rather complicated and not amenable to an analytic treatment in general. It turns out though that its temperature dependence, at low temperature, is controlled by its non-analytic behavior in the following sense. Applying Poisson resummation to (\ref{Z4dKK}) gives
\be
\label{Z4dthermal1}
\log Z_{\rm 4d} = \sum_{l = -\infty}^\infty \beta L \int {dk \over 2\pi} {d\omega \over 2\pi}
e^{-i \beta \omega l} \log Z_{\rm 2d}\big(\mu = \sqrt{k^2 + \omega^2}\big)
\ee
The $l = 0$ term is proportional to $\beta$ and it gives the Casimir
energy at zero temperature. Thermal corrections to this are given by
\be
\label{Z4dthermal}
\log Z_{\rm 4d,T} = {\beta L \over \pi} \sum_{l = 1}^\infty \int_0^\infty \mu d\mu J_0(\beta l \mu)
\log Z_{\rm 2d}(\mu)
\ee
where we set $\omega = \mu \cos \theta$, $k = \mu \sin \theta$ and integrated over $\theta$. It is clear that the behavior of (\ref{Z4dthermal}) at low temperature, $\beta \rightarrow \infty$, is related to the
behavior of $\log Z_{\rm 2d}$ as $\mu \rightarrow 0$.  If $\log Z_{\rm 2d}(\mu)$ is analytic in $\mu^2$, then the 4d free energy vanishes exponentially at low temperature, which is the case if the parameters $a, b_1, b_2$ in Fig. 1 are held fixed and finite. The difference in the non-analytic behavior of $\log Z_{\rm 2d}$ for the three special plate configurations we have studied, results to interesting variations in the temperature dependence as we shall show below.

\vskip .15in
\noindent
{\it Plate with a slit}

The bulk contribution from the regions above and below the middle plate is that of a 4d ideal Bose gas. The renormalized thermal free energy after we subtract the free energy of the ``big box" and the self-energy of the middle plate is 
\be
F^{\rm bulk}_{\perp,T} = {\zeta(3) \over{4 \pi}}L w T^3
\ee
The direct contribution to the free energy is related to the free energy of a 3d ideal gas occupying the region corresponding to the gap and it is exponentially suppressed. At low temperatures, $aT \ll1$, the thermal wavelength is larger than the size of the gap and this leads to exponential suppression. 

The non-analytic behavior of the first diffractive contribution to $\log Z_{2d}$ is of the form
\be
-\log Z_{2d} = - {{7 \zeta(3)} \over {\pi^4}} (\mu a)^2 \log \mu a + ((\mu a)^4 \log \mu a \rm ~terms~+~\cdots) 
\ee
This gives rise to a contribution to thermal free energy proportional to $T^4$. In particular 
\be
F^{(1),\rm diffr}_{\rm slit, T} = - {{7 \zeta(3)} \over {90 \pi}} Lw^2T^4 + {\cal O}(T^6)
\ee
The total thermal free energy up to this order is
\be
F_{\rm slit, T} = {\zeta(3) \over {4 \pi}} LwT^3 - {{7 \zeta(3)} \over {90 \pi}} Lw^2T^4 + {\cal O}(T^6)
\ee
The leading contribution is coming from the bulk term and can be thought of as an excluded area effect.  The diffractive contribution is subleading at low temperatures. The higher order diffractive terms will change the overall coefficient of $T^4$, but a rough estimate shows a change of less than $10 \%$.  
\vskip .15in
\noindent
{\it Perpendicular plates}

Similar results hold for perpendicular plates. The $a$-dependent part of the bulk and direct contributions are the same as in the case of the plate with a slit. The main difference is in the non-analytic behavior of $\log Z_{2d}$, which is
\be
-\log Z_{2d} = {{ \zeta(3)} \over {4 \pi^4}} (\mu a)^4 \log \mu a + ((\mu a)^6 \log \mu a ~\rm terms~+~\cdots) 
\ee
which results in a thermal free energy contribution proportional to $T^6$. 
\be
F^{(1),\rm diffr}_{\perp,T} = - {{16 \pi \zeta(3)} \over {945}} La^4 T^6 + {\cal O}(T^8)
\ee
The total thermal free energy up to this order is
\be
F_{\perp, T} = -{\zeta(2) \over {8 \pi}} LT^2 + {\zeta(3) \over {4 \pi}} LaT^3 - {{16 \pi  \zeta(3)} \over {945}} L a^4 T^6 + {\cal O}(T^8)
\ee
The first two terms arise from the bulk determinants and can be thought of as an excluded area and perimeter effect. They agree with the results on the thermal Casimir force found in \cite{KlingGies}, \cite{GiesWeber}. Diffractive effects are subleading $\sim~T^6$ and at low temperatures are substantially weaker than in the case of a plate with a slit.

\vskip .15in
\noindent
{\it Infinite, semi-infinite parallel plates}

The analysis in the case of an infinite, semi-infinite parallel plates is slightly more involved but along the same lines. We find that the thermal free energy can be decomposed into an excluded volume contribution 
\be
F_{||,T}^{\rm ex} = {\zeta(4) \over \pi^2} \, V_{\rm ex} T^4 - {\zeta(3) \over 8 \pi} A_{\rm ex} T^3 + {\zeta(2) \over 16 \pi} P_{\rm ex} T^2
\ee
and a diffractive edge contribution
\be
\label{EdgeConclusion}
F^{\rm edge}_{||,T} = - {2 \zeta(4) \over \pi^3} \, (bT)^4 \left(\log(2bT) + {\zeta'(4) \over \zeta(4)}\right) {L \over{b^2}}+ {{3 \zeta(5)} \over {4 \pi}} L b^3T^5 + \cdots 
\ee
$V_{\rm ex},~A_{\rm ex},~P_{\rm ex}$ is the excluded volume, area and perimeter of the region between the two plates. The appearance of these geometric terms has to do with the fact that at low temperatures thermal excitations are excluded from the region between the plates.
 
The edge contribution to the thermal Casimir energy in the case of an infinite/semi-infinite parallel plates was studied by Gies and Weber in \cite{GiesWeber} using the worldline formalism. They found that their data was well fit, at low temperatures, in terms of a power-law temperature dependence with a non-integer exponent $\sim T^{3.74}$. Our analysis however shows that the
non-integer power law found in \cite{GiesWeber} is actually due to a logarithmic temperature
dependence of the form $T^4 \log T$.

\section{Casimir interaction between holes/slits on a plate\label{sec:holes}}

An interesting new result we have derived using the approach outlined here is the Casimir interaction between holes or slits on a Dirichlet plate \cite{Kabat2}. The relevant geometry for two holes is shown in Fig.2.
\begin{figure}
\begin{center}
\scalebox{.8}{\includegraphics{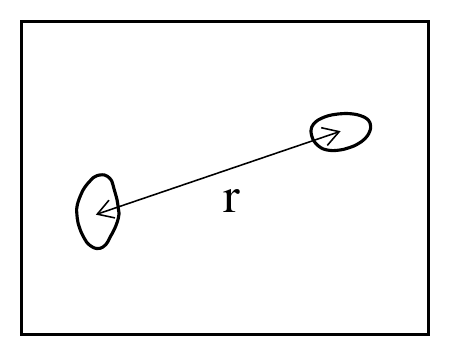}}
\end{center}
\caption{Two holes separated by a distance $r$ on an infinite Dirichlet plate.}
\end{figure}
The functional integral is now 
\be
\label{Zmatrix}
- \log Z = \beta \int_{-\infty}^\infty {d\mu \over 2 \pi} ~{1 \over 2} \, {\rm Tr} \, \log \left(
\begin{array}{cc}
{\cal O}_{11} & {\cal O}_{12} \\
{\cal O}_{21} & {\cal O}_{22}
\end{array}\right)
\ee
where ${\cal O}_{ij} = \langle \rm modes~ on ~hole~ i~|{\cal O}|\rm modes~ on~ hole~ j\rangle$.
For separations large compared to the size of the holes,
\be
{\cal O}_{12}= \langle {\bf x}_1 \vert {\cal O} \vert {\bf x}_2 \rangle  \approx 
\langle 0 \vert \sqrt{-\nabla^2 + \mu^2} \, \vert r \rangle 
 =  - {1 \over 2 \pi r^3} (1 + \mu r) e^{-\mu r} \nonumber
\ee
For $r \gg ({\rm hole ~~size})$ we can expand (\ref{Zmatrix}) in powers of the off-diagonal entries which are small. The first order expansion gives an interaction energy of the form
\bea
\label{Eint}
E_{\rm int} &=& - {1 \over 2\pi} \int_0^\infty d\mu \,
{\rm Tr} \, \left[({\cal O}_{11})^{-1} {\cal O}_{12} ({\cal O}_{22})^{-1} {\cal O}_{21}\right]\nonumber\\
&=&  - {5 \over 32 \pi^3} { Q_1 Q_2\over  r^7}
\eea
where the charge associated with hole i is 
\bea
Q_i &=& \int_{\hbox{\tiny hole $i$}} d^2x \,d^2x' \,\, \langle {\bf x} \vert \left({\cal O}_{ii}\right)^{-1} \vert {\bf x}' \rangle \nonumber \\
&\approx& 1.28 R^3 ~({\rm round}), ~0.228 L^3 ({\rm square})\nonumber
\eea
Eq.(\ref{Eint}) is reminiscent of the van der Waals interaction between atoms. As in that case, the $1/r^7$ dependence is universal at large distances \cite{Kabat2}. Similar results are available for the interaction energy between two far away infinitely long slits on a Dirichlet plane. In this case we find 
\bea
E_{\rm int} &=& - {1 \over 2} \int  {d^2\mu \over {(2 \pi)}^2}\,
{\rm Tr} \, \left[({\cal O}_{11})^{-1} {\cal O}_{12} ({\cal O}_{22})^{-1} {\cal O}_{21}\right]\nonumber\\
 &=&   - {{Q_1 Q_2} \over r^6}
 \label{Eints} 
\eea
where the charge associated with slit i is 
\bea
Q_i & = & \int dx dx' \langle x \vert {\cal O}_{ii}^{-1}\vert x' \rangle \nonumber \\
& = & 2.88 \times 10^{-2} ({\rm slit~width})^2 \nonumber 
\eea
How the interaction energy (\ref{Eint}) depends on distance as the holes approach each other, what is the dynamics of a large number of small mobile holes on a Dirichlet plate and possible experimental observation of such interactions are interesting questions to be explored further.

\section{Conclusions}

In this talk, I have described a general method we developed in calculating the Casimir energy in geometries with apertures and edges. We have found that contributions to Casimir energy due to boundary openings are described in terms of a lower-dimensional, non-local field theory defined on the aperture itself. 

Our method of calculating Casimir energy provides a systematic way of analyzing diffractive contributions in a series expansion, which can be easily generalized to include finite temperature effects, arbitrary dimensions, etc. More work is needed to understand the convergence of the series and justify why this expansion seems to work very well despite the fact that there is no obvious dimensionless parameter which controls such an expansion. It would also be interesting to understand the relation between our expansion scheme and the multiple scattering method developed in \cite{MIT}. 

In all the cases we have analyzed, at zero and at finite temperature, the diffractive effects produce a repulsive Casimir force, opposite to the effect of the bulk and direct contributions. They are subdominant, but their presence diminishes the attractive component of the Casimir force. A general understanding of this and how it relates to the diffractive matrix elements (\ref{OddDiffractive}) and (\ref{EvenDiffractive}) warrants further investigation. 

The dependence of the results I presented above on the curvature of space, intrinsic and extrinsic, as well as the spin of the fields involved are interesting topics to be further explored. 

\ack
It is a pleasure to thank Prof. Cestmir Burdik and his team for the organization of the QTS7 conference. This work was supported by the U.S. National Foundation grant PHY-0758008 and a PSC-CUNY grant.


\section{References}


\begin{thebibliography}{9}


\bibitem{Casimir}
H.B.G. Casimir, 
Proc. K. Ned. Akad. Wet {\bf 51}, 793 (1948);
H.B.G. Casimir and D. Polder, 
\PR~{\bf 73}, 360 (1948).

\bibitem{reviews}
K.A. Milton, 
{\it J. Phys. Conf. Ser.}~ {\bf 161}, 012001 (2009);
K. A. Milton, {\it The Casimir Effect: Physical Manifestations of Zero-Point Energy} (World Scientific, 2001);
M. Bordag, U. Mohideen and V.M. Mostepanenko, 
{\it Phys. Rept.}~ {\bf 353}, 1 (2001);
M. Bordag, G.L. Klimchitskaya, U. Mohideen and V.M. Mostepanenko, {\it Advances in the Casimir Effect}
(International Series of Monographs on Physics, 2009).

\bibitem{Lamoreaux}
S.K. Lamoreaux, 
\PRL~{\bf 78}, 5 (1997); 
{\it Rep. Prog. Phys.}~{\bf 68}, 201 (2005); U. Mohideen and A.Roy, 
 \PRL~{\bf 81}, 4549 (1998).

\bibitem{Gies}
H. Gies and K. Klingm\"uller,  
\PRL~{\bf 96}, 220401 (2006).

\bibitem{KlingGies}
K.~Klingm\"uller and H.~Gies,
{\it J. Phys.} {\bf A41}, 164042 (2008).

\bibitem{GiesWeber}
A.~Weber and H.~Gies,
{\it Phys. Rev.} {\bf D80}, 065033 (2009);
H.~Gies and A.~Weber,
 {\it Int. J. Mod. Phys.} {\bf A25}, 2279 (2010).


\bibitem{Kabat1}
D.~Kabat, D.~Karabali, and V.~P. Nair, 
{\it Phys. Rev.} {\bf D81}, 125013 (2010) [arXiv:1002.3575].

\bibitem{Kabat2}
 D.~Kabat, D.~Karabali, V.~P.~Nair, 
 {\it Phys. Rev.} {\bf D82}, 025014 (2010) [arXiv:1005.4341].


\bibitem{Kabat3}
D. Kabat and D. Karabali, 
{\it Phys. Rev.}~{\bf D84}, 065029 (2011) [arXiv:1107.0952].

\bibitem{MIT}
N.~Graham, A.~Shpunt, T.~Emig, S.~J.~Rahi, R.~L.~Jaffe and M.~Kardar, 
{\it Phys. Rev.} {\bf D81}, 061701 (2010); M. F. Maghrebi, S. J. Rahi, T. Emig, N. Graham, R. L. Jaffe, M. Kardar, 
{\it Proc. Nat. Acad. Sci.} {\bf 108}, 6867 (2011); M. F. Maghrebi, N. Graham, 
{\it Europhys. Lett.} {\bf 95}, 14001 (2011); N. Graham, A. Shpunt, T. Emig, S. J. Rahi, R. L. Jaffe, M. Kardar, 
{\it Phys. Rev.} {\bf D83}, 125007 (2011).






\end{thebibliography}
\end{document}